# Some Problems with Negative Refraction*

## by John Michael Williams

P. O. Box 2697,

Redwood City, CA 94064

`jwill@AstraGate.net`

### *Abstract*

J. B. Pendry's "Negative Refraction Makes a Perfect Lens" is analyzed. It appears that several statements may be understood in terms of lens design but not in terms of fundamental behavior of light.

In a recent Letter, "Negative Refraction Makes a Perfect Lens" (*PRL*, 2000, v. **85**(18), pp. 3966 - 3969), author J. B. Pendry makes several puzzling claims.

First, the author begins by stating that "[lens] limitations are dictated by wave optics: no lens can focus light onto an area smaller than a square wavelength." This is not strictly true. The imaging limitations are given by Heisenberg's Uncertainty Principle, not wave optics. In the one-dimensional form, this limitation states that $4\boldsymbol{p}\Delta x \Delta p \geq h$, which implies that $4\boldsymbol{p}\Delta x \Delta(h\boldsymbol{n}) \geq h$. So, for frequency in Hz, and momentum per *c*, the smallest resolved detail $\Delta x$ either in the object or in an image of it would be given by,

$$\Delta x \geq \boldsymbol{l}/2, \tag{1}$$

within some factor of what one considered an identifiable $\boldsymbol{l}$. A limit of $\boldsymbol{l}/4$ would be optimistic. The focussing mechanism is irrelevant to this limit; it can be overcome only by taking advantage of predictability in repeated measurements. Phase correction, for example in semiconductor fabrication, uses small, thin-film shapes to take advantage of predictability and requires that the desired geometry of each image be known before imaging. The optical system in the near field then may be designed so that the actual image is a corrected, "perfect" copy of an object up to, but not beyond, the Heisenberg equality in (1).

---
\* Text of preprint posted at *arXiv* as `physics/0105034`.

There is another, different limitation on the resolution of an optical system, namely the size of the entrance pupil. The aperture holding a lens or silver film would determine this size. A small aperture loses those low Fourier spatial frequencies which transform to supply the highest image detail.

Second, the author states just after Eq. (9) that, when both the permittivity $e$ and the permeability $m$ are negative, one must use the negative square root in $n = \sqrt{em}$. However, algebraically, $n^2 = em \equiv (-e)(-m)$: The negation operator for vectors is distributive, for example in $\mathbf{J} = (k/m_0)\mathbf{E} \times \mathbf{B}$, so it appears that one would choose the solution of $n^2 = em$ depending on the direction of propagation. If only one of $e$ and $m$ were negative, the refractive index would be imaginary, not negative. The evanescent fields, of course, are signed to define the same direction of propagation as the field in the imaging medium; continuity in Maxwell's equations requires this and allows no freedom of choice or control.

Pendry's approach does not seem to consider a complex $n$, nor a complex permittivity nor permeability. The question of a complex index of refraction $N = n(1 - i\mathbf{k})$, with $\mathbf{k}$ an absorption coefficient, always has been a source of inconsistency among authors, as lamented in [1]. It therefore would seem unwise of Pendry to prescribe a negative sign in a field expression without consideration of context. However, the sine being a simple odd function, adopting a negative sign convention to describe an angle in Snell's law of ray optics might be useful to represent a negative refractive index in lens design, for example in correction of aberrations. We would have, $n_1 \sin q_1 = n_2 \sin q_2$, with $(-n_2)\sin q_2 \equiv n_2 \sin(-q_2)$. Although it might seem odd to have a convex lens acting concave, there doesn't seem to be anything impossible about a negative $n$ in the context of Snell's law: The sign simply might be moved by the designer to the sine argument, as when accounting for reflections.

Third, the author makes several statements about the evanescent field, but seems not to distinguish between it and the field of a plane wave. Indeed, if the evanescent field at the border of an aperture could be used to increase the aperture size indefinitely, the resolution might be made to approach the limit $\Delta x$ in (1) above. But, for the energy in the evanescent field to improve the image, the entrance pupil diameter would have to be comparable in linear size to the evanescent field, which would be, say, one wavelength. A wavelength-sized system actually seems to be the system for which Pendry's calculations were made. A system with wavelength-sized aperture would experience significant diffractive losses and hardly could image anything, whether amplified or not. Adding evanescent terms to improve an already poor image merely would represent the effect of increasing aperture size at most by a few more wavelengths and could not be expected to approach the equality in (1) above.

Finally, the author states on p. 3967 that "evanescent waves transport no energy". Perhaps better would be to say that they dissipate no energy; they do no work on the medium. However, like all other regions of the field of a propagating electromagnetic wave, the evanescent fields store energy transiently and return it to the rest of the field according to the propagation geometry and the time-course of any associated causality. Coupling of energy by evanescent fields may be demonstrated simply by holding two optic fibers in differentially close contact with one another.